\documentstyle[12pt,epsf]{article}

\def\lsim{\mathrel{\raise.2ex\hbox{$<$}\hskip-.8em\lower.9ex\hbox{$\sim$}}}
\def\gsim{\mathrel{\raise.2ex\hbox{$>$}\hskip-.8em\lower.9ex\hbox{$\sim$}}}

\begin{document}
\thispagestyle{empty}

\renewcommand{\thefootnote}{\fnsymbol{footnote}}

\font\fortssbx=cmssbx10 scaled \magstep2
\hbox to \hsize{
\includegraphics{/NextLibrary/TeX/tex/inputs/uwlogo.ps}
\hskip.35in \raise.1in\hbox{\fortssbx University of Wisconsin - Madison}
\hfill$\vcenter{\hbox{\bf MADPH-97-987}
            \hbox{\bf astro-ph/9703004}
            \hbox{January 1997}}$ }

\vspace{.5in}

\begin{center} {\large\bf Very High Energy Phenomena in the Universe:\\[2mm]
 Summary Talk of Moriond 97}\footnote{Talk given at {\it The XXXIInd Rencontre de Moriond, ``Very High Energy Phenomena in the Universe"}, Les Arcs, France (1997).}\\[4mm]
Francis~Halzen\\
{\it Department of Physics, University of Wisconsin, Madison, WI 53706}\\
\end{center}

\vspace{.4in}

{\small\narrower
This is the summary of a week of very informative presentations on new ways to probe the Universe using gravitational detectors, space and ground based gamma ray telescopes, EeV air shower detectors and neutrino telescopes. Gamma ray bursts and active galaxies were hot theoretical themes in the multi-wavelength discussions.\par}

\vspace{.5in}

\def\large{\normalsize}
\def\Large{\normalsize}
\renewcommand{\thesection}{\arabic{section}.}

\section{Introduction}

Efforts are underway to qualitatively improve the instruments that can push astronomy beyond GeV photon energy, to wavelengths smaller than $10^{-14}$cm, and map the sky in neutrinos and EeV cosmic rays as well as gamma rays. New gravitational wave detectors will explore wavelengths much larger than those of radio astronomy. The diffuse cosmic photon flux is shown in Fig.~1 as a function of photon energy and wavelength. The instruments which have collected the data shown in the figure span 60 octaves in photon frequency, 
\begin{figure}[t]
\centering
\hspace{0in}\epsfxsize=5in\epsffile{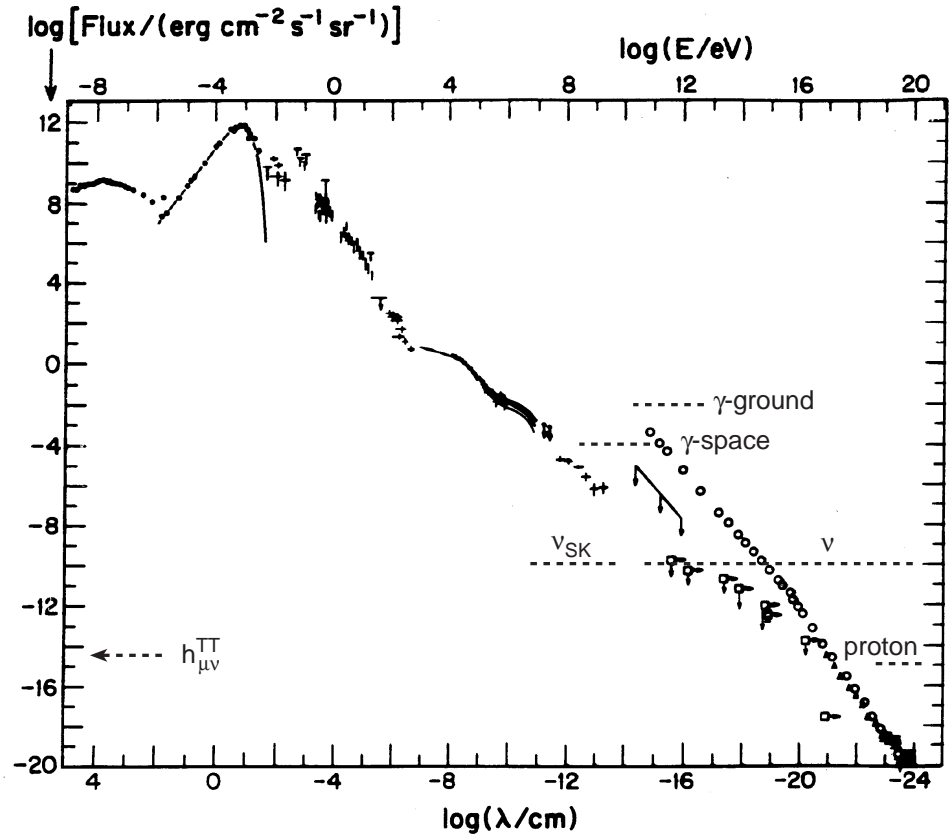}

\caption{Flux of gamma rays as a function of wavelength and photon energy. In the TeV--EeV energy range the anticipated fluxes are dwarfed by the cosmic ray flux which is also shown in the figure.}
\end{figure}
from $10^4$\,cm radio-waves to $10^{-14}$\,cm photons of GeV energy. This is an amazing expansion of the power of  our eyes which scan the sky over less than a single octave just above $10^{-5}$\,cm wavelength. The new astronomy, discussed at this conference, probes the Universe at new wavelengths, smaller than $10^{-14}$\,cm and larger than 1~kilometer. Besides gamma rays, gravitational waves, neutrinos and the very high energy protons which are only weakly deflected by the magnetic field of our own galaxy, become the messengers from the Universe. As exemplified time and again, the development of novel ways of looking into space invariably results in the discovery of  unanticipated phenomena. As is the case with new accelerators, observing the predicted will be slightly disappointing.

\goodbreak
The ``New Astronomy" is sketched in Fig.~1 and covers:
\begin{itemize}
\item
gravitational waves with the commissioning of pairs of resonant bar detectors and LIGO,
\item
gamma ray astronomy, with new satellite-borne detectors expanding from the GeV, into the hundreds of GeV energy region, while, at the same time, second-generation ground-based air Cherenkov telescopes reach down to thresholds of tens of GeV,
\item
cosmic rays which may point at their sources with degree accuracy when their energy exceeds tens of EeV, and
\item
neutrinos, with the first results from Superkamiokande, Baikal and AMANDA presented at this meeting.
\end{itemize}

A novel, but essential aspect of the ``New Astronomy" is that the Universe is not transparent to photons of TeV energy, and above. The same is true for protons once their energy exceeds 50~EeV. In the high energy sky, only neutrinos can reach us from the very edge of the Universe. The transparency of the Universe to photons and protons is shown in Fig.~2. Energetic photons are efficiently decelerated by pair production of electrons on background light above a threshold
\begin{equation}
4E\epsilon \sim (2m_e)^2 \,,
\end{equation}
where $E$ and  $\epsilon$ are the energy of the accelerated and background photon, respectively. Therefore TeV photons are absorbed on infrared light, PeV photons on the cosmic microwave background and EeV photons on radio-waves. It is, for instance, likely that absorption effects explain why Markarian 421, the closest blazar on the EGRET list at a distance of $\sim$100~Mpc, produces the most prominent TeV signal.

\begin{figure}
\centering
\hspace{0in}\epsfxsize=5in\epsffile{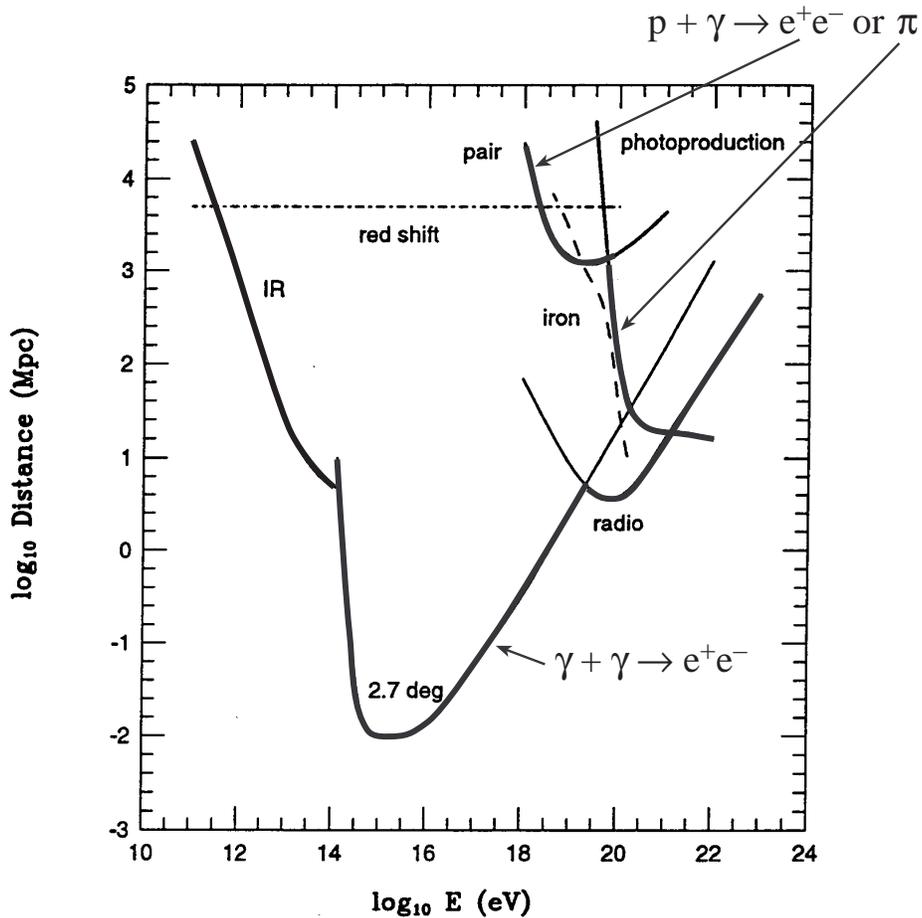}

\caption{Absorption of photons and protons on the interstellar light shown in Fig.~1. Shown is the absorption length in megaparsecs as a function of photon/proton energy. }
\end{figure}

Also protons interact with background light by photoproduction of the $\Delta$-resonance just above the threshold for producing pions:
\begin{equation}
2E_p\epsilon > \left(m_\Delta^2 - m_p^2\right) \,.
\label{eq:threshold}
\end{equation}
The major source of energy loss of $\sim$100~EeV protons is photoproduction of pions on a target of cosmic microwave photons. The Universe is therefore opaque to the highest energy cosmic rays, with an absorption length of only tens of megaparsecs when their energy exceeds $10^{20}$~eV. Lower energy protons, below threshold (\ref{eq:threshold}), do not suffer this fate. They can, however, not be used for astronomy because their directions are randomized in the microgauss magnetic field of our galaxy.

\section{Gravitational Waves\protect\cite{gravity}}

The asymmetric collapse, e.g.\ of a rotating star, near the center of our galaxy will result into the supernova display astronomy is waiting for: the simultaneous observation of light, neutrinos and gravitational waves could be the scientific event of all times. If we make the optimistic assumption that a similar amount of energy is emitted in gravitational waves and in light, i.e.\ one hundreds of a solar mass, gravitational antennas will detect a whopping signal of $\delta h=10^{-18}$. This deformation of the transverse components of the space-time tensor $h^{TT}_{\mu\nu}(x-ct)$ is detected at Earth in the form of gravitational waves. The passage of gravitational waves is revealed by a change in distance $L$ between a pair of masses which are, ideally, separated by half the wavelength of the signal:
\begin{equation}
\delta h = {\delta L\over {L\over2}} \,.
\end{equation}
Given the quadrupole structure of the waves, sensitivity is improved by using an orthogonal pair of spectrometers; see Fig.~3.

\begin{figure}[t]
\centering
\hspace{0in}\epsfxsize=4in\epsffile{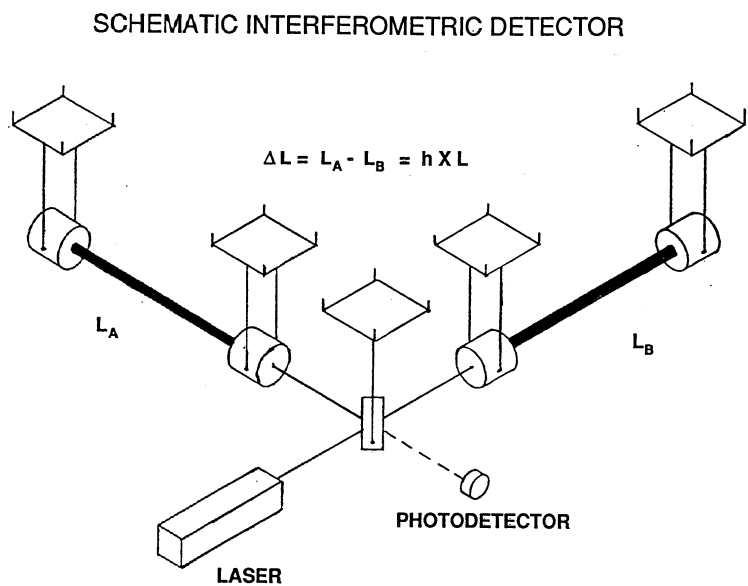}

\caption{Schematic of LIGO.}
\end{figure}

Besides gravitational collapse, other anticipated sources of gravitational waves come in two categories: point sources and stochastic radiation. In the first category we identify the coalescence of a pair of neutron stars or black holes, the spin-down radiation of pulsars and binary systems. Examples of stochastic radiation include gravitational waves emitted by early topological defects (today such radiation is strongly limited by the fact that it would destroy pulsar timing if present at significant levels) and the universal background of gravitons. Gravitons couple more weakly than neutrinos. In standard big bang cosmology they decouple at $10^{-22}$ seconds and, in contrast with the $2K$ cosmic neutrinos, we can contemplate realistic methods for their detection. Any trace of gravitons would, of course, be wiped out during an epoch of inflation.

Because interferometers operate by comparing candidate signals to templates of predicted signatures of gravitational waves, the theoretical computation of the deformation of the metric tensor is critical. Figure~4 displays the gravitational wave from a binary system calculated to leading order in the potential
\begin{equation}
\epsilon = {G\over c^2} \, {m\over r} = \left(v\over c\right)^2 \,.
\end{equation}
Such post-Newtonian calculations have been performed to order $\epsilon^{5/2}$ and results to third order will soon be available.

\begin{figure}
\centering
\hspace{0in}\epsfxsize=5in\epsffile{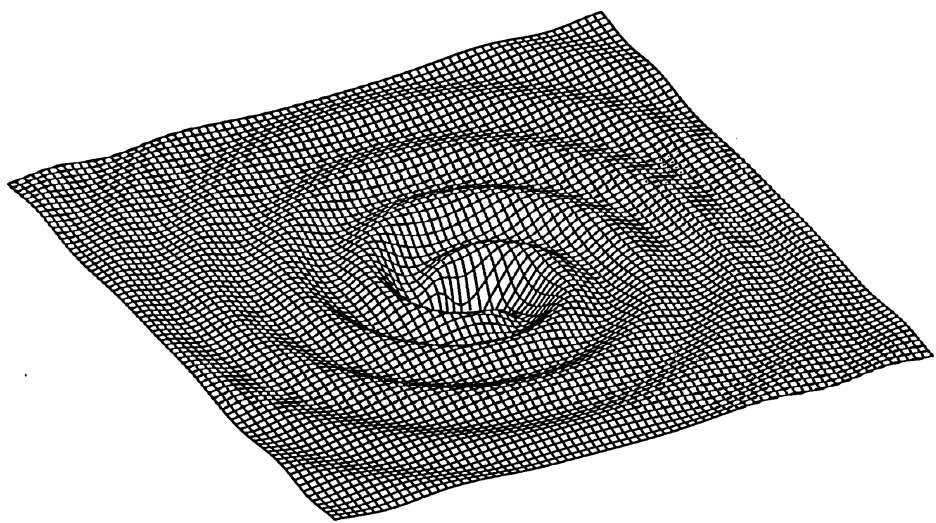}

\caption{Gravitational wave from a binary.}
\end{figure}

\goodbreak
In the push to commission improved instruments we can identify three directions: i)~coincident resonant bar detectors, ii)~LIGO with a sensitivity in the $10{\sim}10^3$~Hz frequency range, and iii)~space-based interferometers such as VIRGO. Because time varying gravitational fields such as clouds, prevent the operation of ground-based detectors at frequencies below 1~Hz, space instruments are our only window to possible signals from supermassive objects such as the black holes which power active galaxies.

LIGO construction has passed its two thirds mark towards completion, with progress measured in terms of dollars. With $L = 4$~km in Eq.~(3), the instrument will initially reach a sensitivity of $\delta h \geq 10^{-21}$. The challenge is to scale up the prototype, presently operating at Caltech, by a factor of 100 without loss of sensitivity. Optical precision has to reach one thousands of the micrometer wavelength of the laser light monitoring the distance between the pairs of masses in Fig.~3. Although a pair of neutron stars coalesces every minute, somewhere in the Universe, the signal of a nearby event is expected to be only of order 1 per year. The sensitivity of the instrument is limited by its frequency range to systems not much heavier than a solar mass. Yet black hole mergers may be viewed by LIGO out to $10{\sim}10^2$~Mpc, at a rate of 10 per year, possibly more.

\section{Gamma Ray Astronomy on Earth and in Space\protect\cite{gamma}}

After two decades, ground-based gamma ray astronomy has become a mature science. The shower imaging method, pioneered by the Whipple telescope, emerged as the winning technique. Data taken on May 7, 1996 on the flaring active galaxy Markarian 421 testify to this statement. Shown in Fig.~5 is the spectacular enhancement of photons in the direction of the source, over a small uniform background of sky light. At this meeting, appropriately, the Whipple group presented a string of interesting new results which include:

\begin{itemize}

\item
evidence that a source can be tracked to large zenith angle. This raises the threshold of the instrument and provides evidence that the Markarian 421 spectrum extends well beyond 5~TeV,
\item 
evidence that this blazar emits TeV-photons in bursts with a duration of order a few days; see Fig.~6,
\item
first hints of a correlation between the optical and TeV variability of blazar jets,
\item
observation of a burst lasting only 15 minutes, suggesting emission from very localized regions of the galaxy, presumably the jet (more about jets later on),
\item
identification of the first TeV source not detected in the GeV region by satellite-borne experiments such as EGRET: Markarian 501, another nearby blazar.

\end{itemize}

\begin{figure}[t]
\centering
\hspace{0in}\epsfxsize=2.5in\epsffile{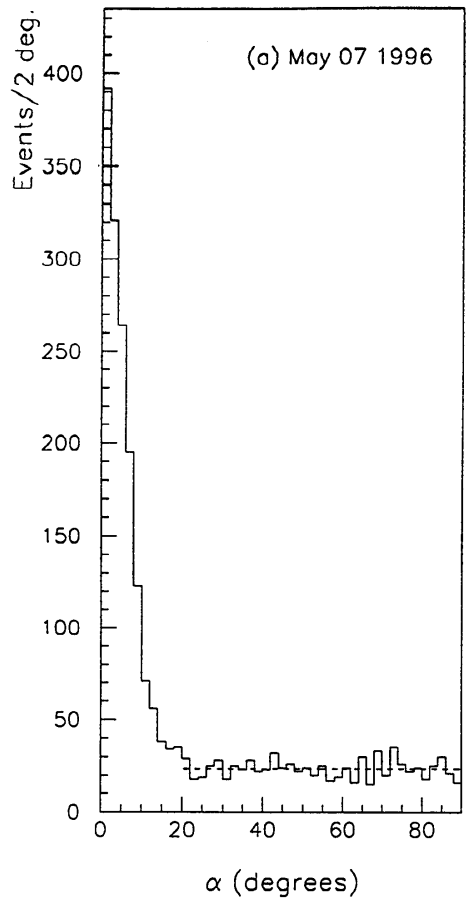}

\caption{Signal and background of TeV photons during a flare of  Markarian 421.}
\end{figure}

\begin{figure}
\centering
\hspace{0in}\epsfxsize=5in\epsffile{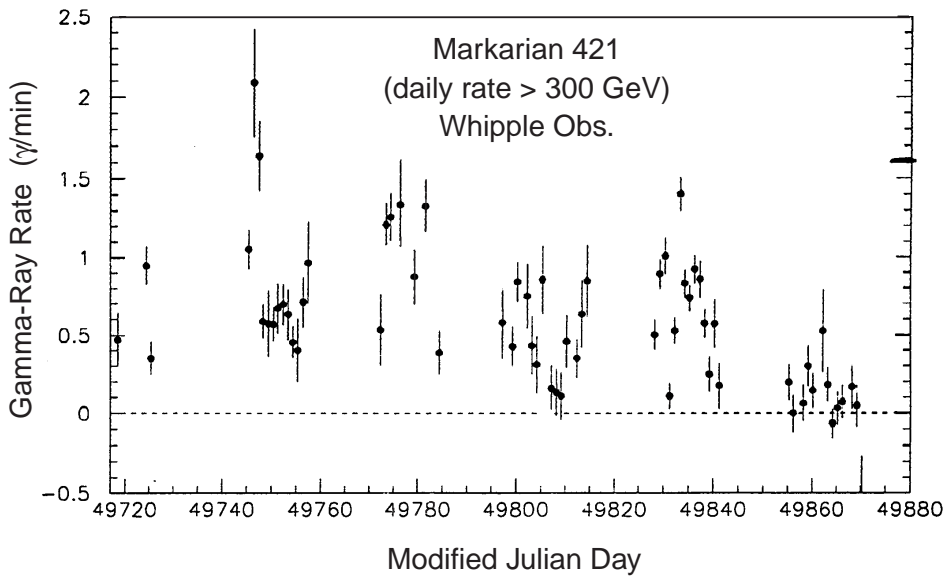}

\caption{ Time variation of the TeV photon flux from the blazar Markarian 421.}
\end{figure}

Then there was the dog that didn't bark: no TeV emission from the supernova remnants IC433 and $\gamma$-Cygni, only upper limits from the Whipple and HEGRA detectors. If old supernova remnants are indeed the accelerators of the bulk of the cosmic rays, a healthy TeV signal is expected from the production and decay of neutral pions produced in the nuclear interactions of the nuclei with ambient matter in the shock. Such systems are a test-bed for models of shock acceleration and these observations did not go unnoticed. Models have already been tweaked in order to accommodate the upper limits in the TeV region. The conclusion is inevitably that accelerated electrons which scatter ambient light to GeV energy, are the origin of most of the flux observed by EGRET. Where do the cosmic rays come from? They are, most likely, only produced in the very late stages of the remnant, or in the rare supernova remnants with strong winds.

The field of gamma ray astronomy is buzzing with activity to construct second-generation instruments. Space-based detectors are extending their reach from GeV to TeV energy with AMS and, especially, GLAST, while the ground-based Cherenkov telescopes are designing instruments with lower thresholds. In the not so far future both techniques should generate overlapping measurements in the $10 \sim 10^2$~GeV energy range; see Fig.~7. While all ground-based experiments aim at lower threshold, better angular- and energy resolution, and a longer duty cycle, one can identify a three-prong attack on the construction of improved air Cherenkov telescopes:

\renewcommand{\theenumi}{{\it\roman{enumi}}}
\begin{enumerate}

\item
larger mirror area, exploiting the parasitic use of solar collectors during nighttime (CELESTE and STACEY),
\item
better, or rather, ultimate imaging in the 17~m MAGIC mirror,
\item
larger field of view using multiple telescopes (VERITAS, HEGRA and TOKYO telescope arrays).

\end{enumerate}

\begin{figure}[t]
\centering
\hspace{0in}\epsfxsize=3.5in\epsffile{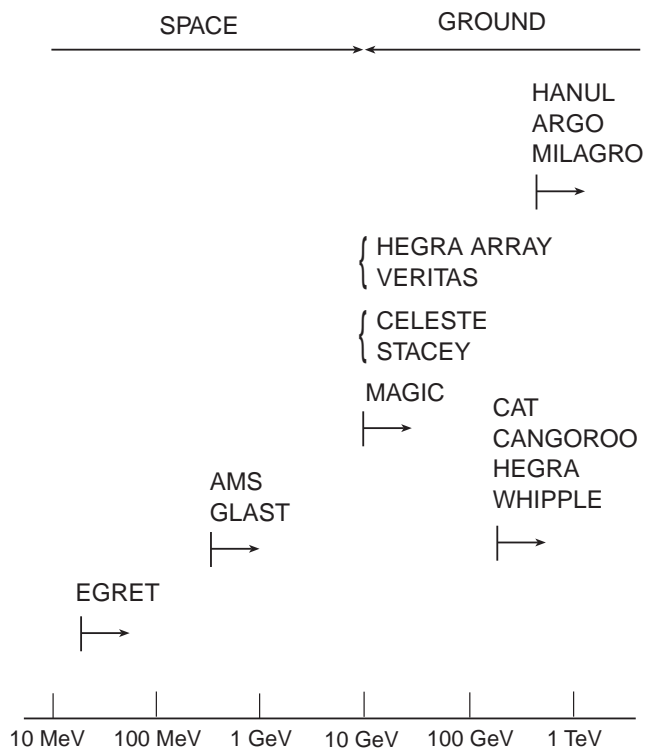}

\caption{Thresholds of gamma ray telescopes.}
\end{figure}

VERITAS, for example, is an array of 9 upgraded Whipple telescopes, each with a field of view of 6 degrees. These can be operated in coincidence, or be pointed at 9 different 6 degree bins in the night sky, thus achieving a very large field of view. The merits of stereo imaging, i.e.\ viewing the same air shower with multiple, imaging telescopes, was hotly debated. Encouraging initial measurements were presented by the TOKYO array, located in Utah, and by the HEGRA telescope array. The debate will be settled by experiment.

There is a dark horse in this race: Milagro. The Milagro idea is to lower the threshold of conventional air shower arrays to 100~GeV by uniformly instrumenting an area of $10^3$~m$^2$ or more (no sampling!). For time-varying signals, such as bursts, the threshold could be even lower. One can instrument a pond with photomultipliers (Milagro), or cover a large area with resistive plate chambers (ARGO), or even with muon detectors (Hanul) which identify point sources of muons produced in photon-induced air showers.

The conference was buzzing with rumors of observations of photons from active galaxies with energies as high as 50~TeV. Confirmation would revolutionize high energy astrophysics; we will return to this in the discussion of active galaxies.

\section{Proton Astronomy: EeV Cosmic Rays\protect\cite{cosmic}}

Around 1930 Rossi and collaborators discovered that the bulk of the cosmic radiation is not made up of gamma rays. This marked the beginning of what was then called ``the new astronomy'', and we refer to as cosmic ray physics today. It is only ``astronomy" above $5 \times 10^{19}$~eV or so, where the arrival directions of the charged cosmic rays are not scrambled by the ambient magnetic field of our own galaxy.  As already mentioned, we suspect that the bulk of the cosmic rays are accelerated in the blastwaves of supernovae exploding into the interstellar medium. This mechanism has the potential to accelerate particles up to energies of $10^3$~TeV where the cosmic ray spectrum suddenly steepens: the ``knee" in the energy spectrum. We have no clue where and how cosmic rays with energies in excess of $10^3$~TeV are accelerated. We are not even sure whether they are protons or iron, or anything else. The origin of cosmic rays with energy beyond the ``knee" is one of the oldest unresolved puzzles in science.

It is sensible to assume that, in order to accelerate a proton to energy $E$, the size $R$ of the accelerator must be larger than the gyroradius of the particle in the accelerating field $B$:
\begin{equation}
R > R_{\rm gyro} = {E\over B} \,.
\end{equation}
This yields a maximum energy
\begin{equation}
E < BR 
\end{equation}
by dimensional analysis and nothing more. The largest structures in our own galaxy are old supernova remnants with $R \simeq 10^2$~parsecs. For the microgauss field of the galaxy $B \simeq 10^{-6}$~G, equation (5) yields a maximum energy of $10^5$~TeV. Inefficiencies in the supernova shock acceleration mechanism will reduce this upper limit by a factor $v_{\rm shock} / c \simeq 0.1$, as well as by another penalty factor for converting shock energy into acceleration, which is also of order 0.1. Therefore, the highest energy one can reach in practice is $10^3$~TeV, i.e.\ the energy of the ``knee" in the cosmic ray spectrum.

Cosmic rays with energy in excess of $10^{20}$~eV have been observed, some five orders of magnitude in energy above the supernova cutoff. To reach higher energy, one has to dramatically increase $B$ and/or~$R$. This argument is difficult to beat --- it is basically dimensional. Although imaginative arguments actually do exist to avoid this impasse, it is generally believed that our galaxy is too small and its magnetic field too weak to accelerate the highest energy cosmic rays. Nearby active galactic nuclei distant by $\sim100$~Mpc are obvious extra-galactic candidates. The jets of blazars support magnetic fields of 10~G over distances of $10^{-2}$~parsecs or more. Using Eq.~(5) we reach energies of $10^{20}$~eV, possibly higher because of beaming. Such speculations are reinforced by the fact that a cursory glance at the EGRET and Whipple results is sufficient to convince oneself that blazars are also the dominant (exclusive?) sources of the highest energy gamma rays.

This raises the obvious question whether the highest energy cosmic rays point back to blazars? Astronomy with protons becomes possible once their energy has reached a value where their gyroradius in the microgauss galactic field exceeds the dimensions of the galaxy. From this point of view protons with $10^{20}$~eV energy point at their sources with degree-accuracy. At this energy, their mean-free-path in the cosmic microwave background is unfortunately reduced to only tens of megaparsecs; see Fig.~2. A clear window of opportunity emerges: Are the directions of the cosmic rays with energy in excess of  $\sim 5 \times 10^{19}$~eV correlated to the nearest AGN (red-shift $z$ less than 0.02), which are known to be clustered in the so-called ``super-galactic" plane? Although far from conclusive, there is some evidence that such a correlation may exist, but not all experiments agree. Lack of statistics at the highest energies is a major problem.

Another problem is that the pointing accuracy is not really understood. It depends on the distance $d$ to the source and the gyroradius in the intergalactic magnetic field:
\begin{equation}
\theta \cong {d\over R_{\rm gyro}} = {dB\over E} \,,
\end{equation}
or scaled to units relevant to the problem
\begin{equation}
{\theta\over0.1^\circ} \cong { \left( d\over 1{\rm\ Mpc} \right) \left( B\over 10^{-9}{\rm\,G} \right) \over \left( E\over 3\times10^{20}\rm\, eV\right) }\,.
\end{equation}
Speculations on the strength of the inter-galactic magnetic field range from $10^{-7}$ to $10^{-12}$~Gauss. For a distance of 100~Mpc, the resolution may therefore be anywhere from sub-degree to nonexistent. It is reasonable to expect that magnetic fields are higher in regions where matter is clustered. Higher values may therefore be appropriate for the local cluster and the super-galactic plane, regions relevant to this problem. Optimistically, one may anticipate that future high statistics experiment such as HIRES in Utah and the Auger giant air shower array will provide indirect information on the magnitude of the magnetic fields between galaxies.

Before turning to experiment, a word about topological defects. Figure~8 shows the Fly's Eye cosmic ray spectrum in the EeV energy range, with the highest energy event ostensibly above a visual extrapolation of the lower energy events. One event has been sufficient to claim evidence for a top-down spectrum of ``the" highest energy cosmic ray. It is suggested that the particle is the decay product of a $10^{24}$~GeV (the GUT unification scale) topological defect such as a monopole, string.... This enthusiastic leap of faith can be excused. Topological structures are deeply connected to gauge theories and cannot be studied in accelerator experiments. Non-accelerator particle physics provides unique opportunities here. A topological defect will suffer a chain decay into GUT particles X,Y, which subsequently decay to the familiar weak bosons, leptons and quark-gluon jets. Cosmic ray protons are the fragmentation products of jets. We know from accelerator data that, among the fragmentation products of jets, neutral pions (decaying into photons) dominate protons by two orders of magnitude. Therefore, if topological defects are the origin of  the highest energy cosmic rays, they must be photons. This is a problem because the highest energy event in Fig.~8 is not likely to be a photon. A photon of $3 \times 10^{20}$~eV interacts with the magnetic field of the earth far above the atmosphere and disintegrates into lower energy cascades; on average eight at this particular energy. The measured shower profile of the event does not support this assumption. One can live and die by a single event!

\begin{figure}[t]
\centering
\hspace{0in}\epsfxsize=3.8in\epsffile{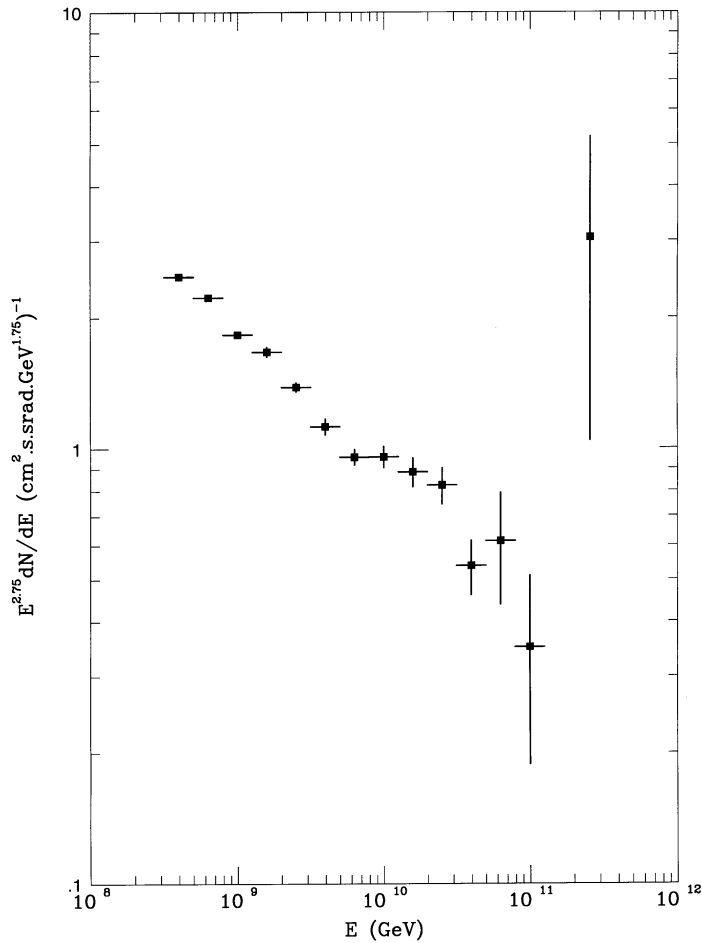}

\caption{The highest energy cosmic ray flux measured by the Utah Fly's Eye.}
\end{figure}

Nevertheless consideration of topological defects deserves our attention because they can only be probed in non-accelerator experiments. That GUT theories are supersymmetric (and so are half of the particles!) has been routinely forgotten in the literature. SUSY in the sky? Although supersymmetry modifies lots of factors in calculations that can, at best, be qualitatively tested, I actually do not know of any clear SUSY signature in a cosmic ray detector. The stable, lightest LSP (the neutralino?) would be very much like a cosmic neutrino and leave no signature.

Fig.~2 should remind us that, if the sources of cosmic rays are beyond $10^2$~Mpc, conventional astronomy cannot find them. In this case, the absorption of the beam on the microwave background becomes a signature for distant sources, irrespective of the pointing precision of the cosmic rays at earth. The main problem today is statistics. After particles with energies in the vicinity of 100~EeV were discovered at Haverah Park, we accumulated a total of four events whose energy clearly exceed $10^{20}$~eV, using three different detectors: AGASA, Yakutsk and the Fly's Eye. The latter is being replaced by a technically superior instrument with larger collection area: the HIRES detector. Construction of a $10^4$~km$^2$ array, one hundred times larger than the AGASA array operating in Japan, has been proposed and will hopefully be launched soon as the ``Auger" project.

\section{Intermezzo: Gamma Ray Bursts\protect\cite{GRB}}

The observation of gamma ray bursts confronts us with a seemingly straightforward problem:

\begin{itemize}
\item
a large amount of energy is released (of order $10^{51}$~ergs, like a supernova),
\item
in high energy photons (and {\bf only} high energy photons --- KeV and above),
\item
in a very short time (typically seconds, or less).
\end{itemize}

One burst a day is, on average, detected from a direction not correlated to our galaxy. All evidence points at a cosmological origin of the sources, with an average redshift $\left<z\right> \sim 1$.

Both the energetics and frequency point at neutron stars: a pair of neutron stars will merge once every minute, somewhere in the Universe. The rest is special relativity. High energy and short time can be accommodated if emission takes place in a site boosted towards the observer by a Lorentz factor $\gamma$. If the source moves towards us we expect that
\begin{equation}
E = E_0 \gamma^n
\end{equation}
and
\begin{equation}
\Delta t = Rc/\gamma^n \,.
\end{equation}
Here $E_0$ is the energy in the frame of the source and $R$ is its size. The power $n$ is model-dependent and different for relativistic shocks in fireballs, jets, oscillating cosmic strings... For instance, models with $n = 2$ require $\gamma = 10^2 \sim 10^3$ in order to describe the qualitative features of gamma ray bursts.

As is the case for most acceleration problems in astrophysics, the devil is in the details. It has not been easy to produce the blueprint for a gamma ray burst. Although the details can be complex, the overall idea of fireball models is that a solar mass of energy is released in a compact region of radius $R\simeq 10^2$\,km. Only neutrinos escape because the fireball is opaque to photons. A significant fraction of the photons in the fireball is indeed above pair production threshold and produces electrons. It is straightforward to show that the optical depth of the fireball is of order $10^{13}$. It is theorized that a relativistic shock, with $\gamma \simeq 10^2$ or more, expands into the interstellar medium and photons escape only when the optical depth of the shock has been sufficiently reduced. There is, at least, one major problem: the shock is expected to lose its energy to the ubiquitous protons in the interstellar medium and fizzle long before it has the opportunity to produce the high energy gamma ray display.

Personally, I find all models unconvincing because they do not naturally explain why the temporal structure of the spectrum of a gamma ray bursts is chaotic and totally different from burst to burst. At this meeting Dar presented a model overcoming this objection.

Dar postulates the formation of a pair of jets along the rotation axis of the merging neutron star pair, a configuration reminiscent of blazar jets. The jet material is presumably composed of protons and heavier nuclei, up to iron, just like cosmic rays. A high Z atom moving with a $\gamma$-factor of order $10^3$ will be photoexcited and ionized by the ambient starlight it encounters in the vicinity of the neutron star merger. From the atom's point of view, eV starlight is indeed boosted to KeV energy. KeV photons, emitted when the atoms relax, are detected as MeV gammas in the observer frame as a result of beaming. Thus two successive boosts turn starlight into MeV gammas according to Eq.~(9), with $n=2$. The baryons, which are a problem in conventional models, are now the solution! More importantly, because light is emitted whenever the shock encounters stars, the detected time profile of the burst represents a tomographic image of the galaxy in the vicinity of the neutron star merger. This explains the lack of organized structure; every burst is different. Monte Carlo simulations based on the distribution of stars near the center of our own galaxy produce bursts with a rich structure, indistinguishable from real data. Si non e vero, e ben trovato... Dar has not offered a detailed model for the propagation of the jet, the astrophysicists were skeptical.

\section{Intermezzo: Blazars\protect\cite{blazar}}

EGRET and the air Cherenkov telescopes have put blazars at the focus of high energy astronomy. EGRET has detected high energy gamma ray emission, in the range 20~MeV--30~GeV, from over 100 sources. Of these sources 16 have been tentatively, and 42 solidly identified with radio counterparts. All belong to the ``blazar" subclass, mostly Flat Spectrum Radio Quasars, while the rest are BL-Lac objects. In a unified scheme of AGN, they correspond to Radio Loud AGN viewed from a position illuminated by the cone of a relativistic jet. Moreover, of the five TeV gamma-ray emitters identified by the air Cherenkov technique, three are extra-galactic and are also nearby BL-Lac objects. The data therefore strongly suggests that the highest energy photons originate in jets beamed at the observer. Several of the sources observed by EGRET have shown strong variability, by a factor of 2 or so over a time scale of several days. Time variability is more spectacular at higher energies; see Fig.~6.

Does pion photoproduction by accelerated protons play a central role in blazar jets? This question was hotly debated. If protons are accelerated along with electrons, they will acquire higher energies, reaching PeV--EeV energy because of reduced energy losses. High energy photons result from photoproduction of neutral pions by protons on the abundant UV photons in the jet. Accelerated protons thus initiate a cascade which dictates the features of the spectrum at lower energy. From a theorist's point of view the proton blazar has attractive features because protons, unlike electrons, efficiently transfer energy from the black hole in the presence of the high magnetic fields required to explain the confinement of the jets. The issue of proton acceleration can be settled experimentally because the proton blazar is a source of high energy protons and neutrinos, not just gamma rays.

First order Fermi acceleration offers a very attractive model for acceleration in jets, providing, on average, the right power and spectral shape. Confronted with the challenge of explaining a relatively flat multi-wavelength photon emission spectrum which extends to TeV energy, models have converged on the blazar blueprint shown in Fig.~9. Particles are accelerated by Fermi shocks in bunches of matter travelling along the jet with a bulk Lorentz factor of order $\gamma \sim 10$. Ultra-relativistic beaming with this Lorentz factor provides the natural interpretation of the observed superluminal speeds of radio structures in the jet. In order to accommodate bursts lasting a day in the observer's frame, the bunch size must be of order $\Gamma c \Delta t \sim 10^{-2}$~parsecs. Here $\Gamma$ is the doppler factor, which for observation angles close to the jet direction is of the same order as $\gamma$. These bunches are, in fact, more like sheets, thinner than the jet's width of roughly 1~parsec. The observed radiation at all wavelengths is produced by the interaction of the accelerated particles in the sheets with the ambient radiation in the AGN, which has a significant component concentrated in the so-called ``UV-bump\rlap".

\begin{figure}[h]
\centering
\epsfxsize=3.25in\hspace{0in}\epsffile{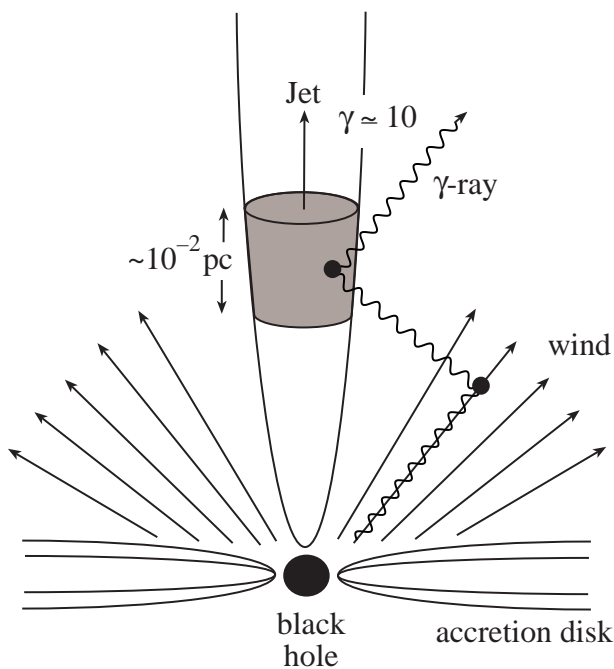}

\caption{Possible blueprint for the production of high energy photons and neutrinos near the super-massive black hole powering an AGN. Particles, accelerated in sheet like bunches moving along the jet, interact with photons, radiated by the accretion disk or, produced by the interaction of the accelerated particles with the magnetic field of the jet.}
\end{figure}

In electron models the multi-wavelength spectrum consists of  three components: synchrotron radiation produced by the electron beam on the $B$-field in the jet, synchrotron photons Compton scattered to high energy by the electron beam and, finally, UV photons Compton scattered by the electron beam to produce the highest energy photons in the spectrum. The picture has a variety of problems. In order to reproduce the observed high energy luminosity, the accelerating bunches have to be positioned very close to the black hole. The photon target density is otherwise insufficient for inverse Compton scattering to produce the observed flux. This is a balancing act, because the same dense target will efficiently absorb the high energy photons by $\gamma\gamma$ collisions. The balance is difficult to arrange, especially in light of observations showing that the high energy photon flux extends beyond TeV energy. The natural cutoff occurs in the 10--100~GeV region. Finally, in order to prevent the electrons from losing too much energy before producing the high energy photons, the magnetic field in the jet has to be artificially adjusted to less than 10\% of what is expected from equipartition with the radiation density. 

For these, and other reasons already mentioned, the proton blazar has been developed. In this model protons as well as electrons are accelerated. Because of reduced energy loss, protons can produce the high energy radiation further from the black hole. The more favorable production-absorption balance far from the black hole makes it relatively easy to extend the high energy photon spectrum above 10 TeV energy, even with bulk Lorentz factors that are significantly smaller than in the inverse Compton models. Because the seed density of photons is still much higher than that of target protons, the high energy cascade is initiated by the photoproduction of neutral pions by accelerated protons on ambient light via the $\Delta$ resonance.

Model-independent evidence that AGN are indeed cosmic proton accelerators can be obtained by observing high energy neutrinos from the decay of charged pions, photoproduced along with the neutral ones. The expected neutrino flux can be estimated in six easy steps.

\begin{enumerate}

\item

The size of the accelerator $R$ is determined by the duration, of order 1 day, over which the high energy radiation is emitted:
\begin{equation}
R=\Gamma t c = 10^{-2}\mbox{ parsecs for }t = 1\mbox{ day and }\Gamma = 10.
\end{equation}

\item

The magnitude of the $B$-field can be calculated from equipartition with the electrons, whose energy density is measured experimentally:
\begin{equation}
{B^2 \over 2 \mu_0}  = \rho\rm (electrons) \sim 1\ erg/cm^3.  
\end{equation}
This yields a value for the magnetic field of 5~Gauss.

\item

In shock acceleration the gain in energy occurs gradually as a particle near the shock scatters back and forth across the front gaining energy with each transit. The proton energy is limited by the lifetime of the accelerator and the maximum size of the emitting region, $R$
\begin{equation}
E  < K Z e B R c\,.
\label{eq:Emax}
\end{equation}
Here  $Ze$ is the charge of the particle being accelerated and $B$
the ambient magnetic field. The upper limit basically follows from dimensional analysis. It can also be derived from the simple requirement that the gyroradius of the accelerated particles must be contained within the accelerating region $R$; see Eq.~(5). The numerical constant $K\sim 0.1$ depends on the details of diffusion in the vicinity of the shock, which determine the efficiency by which power in the shock is converted into acceleration of particles. In some cases it can reach values close to 1. The maximum energy reached is
\[
E_{\rm max} = e B R c = 5 \times 10^{19} \rm\ eV
\]
for $B = 5$~Gauss and  $R = 0.02$~parsecs. We here assumed that the boost of the energy in the observer's frame approximately compensates for the efficiency factor, i.e.\ $K \Gamma \sim 1$. 

The neutrino energy is lower by two factors which take into account i) the average momentum carried by the secondary pions relative to the parent proton ($\left< x_F\right> \simeq 0.2$) and ii) the average energy carried by the neutrino in the decay chain $\pi^+ \rightarrow \nu_\mu \mu^+ \rightarrow  e^+ \nu_e \bar{\nu}_\mu$, which is roughly 1/4 of the pion energy because equal amounts of energy are carried by each lepton. The maximum neutrino energy is 
\begin{equation}
E_{\nu\,\rm max} = E_{\rm max} \left< x_F\right> {1\over4} \simeq 10^{18}\rm\, eV\,,
\end{equation}
i.e.\ neutrinos reach energies of $10^3$~PeV.

\item

The neutrino spectrum can now be calculated from the observed gamma ray luminosity. We recall that approximately equal amounts of energy are carried by the four leptons that result from the  decay chain $\pi^+ \rightarrow \nu_\mu \mu^+ \rightarrow e^+ \nu_e \bar{\nu}_\mu$. In addition the cross sections for the processes $p\gamma \rightarrow p \pi^0$ and  $p\gamma \rightarrow n \pi^+ $ at the $\Delta$ resonance are in the approximate ratio of 2:1. Thus 3/4 of the energy lost to photoproduction ends up in the electromagnetic cascade and 1/4 goes to neutrinos, which corresponds to a ratio of neutrino to gamma luminosities ($L_\nu$/$L_\gamma$) of $1/3$. This ratio is somewhat reduced when taking into account that some of the energy of the accelerated protons is lost to direct pair production ($p +\gamma\rightarrow e^+ e^- p$):
\begin{equation}
L_\nu\, =\,\frac{3}{13}L_\gamma\,.
\end{equation}
In order to convert above relation into a neutrino spectrum we have to fix the spectral index. We will assume that the target photon density spectrum is described by a $E^{-(1+\alpha)}$ power law, where $\alpha$ is small for AGN with flat spectra. The number of target photons above photoproduction threshold grows when the proton energy $E_p$ is increased. If the protons are accelerated to a  power law spectrum with spectral index $\gamma = 2$, the threshold effect implies that the spectral index of the secondary neutrino flux is also a power law, but with an index flattened by $(1+\alpha)$ as a result of the increase in target photons at resonance when the proton energy is increased: 
\begin{equation}
{dN_\nu\over dE_{\nu}} = {\cal N} 
\left[
{ E_{\nu} \over E_{\nu\rm\,max} }
\right] ^{-(1-\alpha)}.
\end{equation}
For a flat photon target with $\alpha=0$, the neutrino spectrum will flatten by just one unit giving $E{dN_\nu\over dE} \sim \rm constant$. From Eqs. (15) and (16)
\begin{equation}
\int^{E_{\nu\rm\,max}} E{dN_\nu\over dE_\nu} dE_\nu \simeq 
{\cal N} {E_{\nu\rm\,max}^2 \over 1+\alpha} \simeq {3\over13}L_\gamma \,.
\end{equation}

\item

Assuming that the high energy $\gamma$ ray flux from Markarian 421 results from cascading of the gamma ray luminosity produced by Fermi accelerated protons, we obtain the neutrino flux from the measured value of $L_{\gamma}$ of $2 \times 10^{-10}$~TeV~cm$^{-2}$~s$^{-1}$: 
\begin{equation}
{dN_\nu\over dE_\nu} = {3\over13}{L_\gamma\over E_{\nu\rm\,max}} \;{1+ \alpha \over E_\nu} \left[ { E_{\nu} \over E_{\nu\rm\,max} } \right]^{-\alpha} \sim 
{5 \times 10^{-17}\,{\rm cm^{-2}\,s^{-1}}\over E} \,,
\end{equation}
where the numerical estimate corresponds to $\alpha=0$ and the value of $E_{\nu\rm\,\max}$ of Eq.~(14). The neutrino flux is essentially determined by the value for $E_{\nu\rm\,\max}$. 

\item 

In order to calculate the diffuse flux from the observed blazar distribution, we note that the EGRET collaboration has constructed a luminosity function covering the observation of the $\sim$20 most energetic blazars and estimated the diffuse gamma ray luminosity. From the ratio of the diffuse gamma ray flux and the flux of Markarian 421, we obtain that the effective number of blazars with Markarian 421 flux is ${\sim}130$~sr$^{-1}$. The diffuse neutrino flux is now simply estimated by multiplying the calculated flux for Markarian 421 by this factor.

\end{enumerate}

This concludes our calculation. It illustrates how the proton, unlike the electron, blazar requires no artificially large $\Gamma$ factors and no fine-tuning of parameters. For the proton blazar, radiation and magnetic fields are in equipartition, the maximum energy matches the $BR$ value expected from dimensional analysis and, finally, the size of the bunches is similar to the gyroradius of the highest energy protons. It is not a  challenge to increase gamma ray energies well beyond the TeV energy range. Reasonable variations of the values of magnetic field strength $B$, the efficiency parameter $K$ and the Doppler boost factor $\Gamma$ may allow us to account for the highest energy cosmic rays with $E\sim 3\times10^{20}~$eV. 

The probability to detect a TeV neutrino is roughly $10^{-6}$. It is easily computed from the requirement that, in order to be detected, the neutrino has to interact within a distance of the detector which is shorter than the range of the muon it has produced. Therefore,
\begin{equation}
P_{\nu\to\mu} \simeq {R_\mu\over \lambda_{\rm int}} \simeq A E^n_\nu \,,
\end{equation}
where $R_{\mu}$ is the muon range and $\lambda_{\rm int}$ the neutrino interaction length. For energies below 1~TeV, where both the range and cross section depend linearly on energy, $n=2$. Between TeV and PeV energies $n=0.8$ and $A=10^{-6}$, with $E$ in TeV units. For EeV energies $n=0.47$, $A =10^{-2}$ with $E$ in EeV.  

We are now ready to compute the diffuse neutrino event rate by folding the neutrino spectrum of Eq.~(18) with the detection probability of Eq.~(19). We also multiply by 130~sr$^{-1}$ for the effective number of sources:
\begin{equation}
\phi^\nu = \int^{E_{\nu\,\rm max}} {dN_\nu\over dE_\nu} P_{\nu\to\mu}(E_\nu) dE_\nu \simeq 40\rm\ km^{-2}\,year^{-1}\,sr^{-1} \,.
\end{equation}
which implies a yield of two neutrinos every three days in a kilometer-scale detector, assuming only $2 \pi$ coverage.

We have already drawn attention to the 10~TeV maximum photon energy as the demarkation line between the electron and proton blazars. The ${\sim}10$~GeV cutoff in the inverse Compton model can be pushed to the TeV range in order to accommodate the Whipple data on Markarian 421, but not beyond. Bringing the accelerator closer to the black hole may yield photons in excess of 10~TeV energy --- they have, however, no chance of escaping without energy loss on the dense infrared background at the acceleration site.

HEGRA has been monitoring the 10 closest blazars, including Markarian 421, with its dual telescope systems: the scintillator and naked photomultiplier detector arrays. They announced at a variety of meetings (but not at this one!) that their upper limit on the flux of photons of 50~TeV energy and above, for the aggregate emission from the ten nearest blazars, may actually be a signal. This could provide the first compelling evidence that blazar jets are indeed true proton accelerators.

\section{The Birth of High Energy Neutrino Astronomy\protect\cite{neutrino}}

With the rapidly expanding Baikal and AMANDA detectors producing their first hints of neutrino events, observation of neutrinos from AGN could establish the production of pions and identify the accelerated proton beams as the origin of the highest energy photons and the highest energy cosmic rays. A definite answer may not be known before neutrino telescopes reach kilometer size. The builders should take note that, although smaller neutrino fluxes are predicted than in the generic AGN models of a few years ago, they are all near PeV energy where the detection efficiency is increased and the atmospheric neutrino background negligible. Because of the beaming of the jets, the neutrinos have a flat spectrum peaking near the $10^6$~TeV maximum energy. The actual event rates are, in the end, not very different.

These models strongly favor the construction of neutrino telescopes following a distributed architecture, with large spacings of the optical modules and relatively high threshold: $\sim 100$~meter spacing among strings and $\sim 20$~meter spacings of OMs along a string (OM: pressure vessel containing a conventional photomultiplier tube and, possibly, data acquisition electronics). Models where most neutrinos have very high energy, open up opportunities for alternative techniques such as the radio technique, or the detection of horizontal air showers with giant air shower arrays. Both were extensively discussed at this meeting. Optimists, on the other hand, can find reasons to anticipate the discovery of AGN neutrinos with much smaller telescopes, such as the existing Macro, Baikal and AMANDA detectors. With a sufficiently high proton target density in the acceleration region, much larger fluxes of neutrinos may be produced in a proton-proton cascade. The predicted fluxes are however model-dependent. It is also possible, even likely, that photons do not escape the source or, escape after significant energyloss. Such absorption effects increase the neutrino flux relative to the observed high energy photon flux, also leading to the prediction of larger neutrino fluxes.

The motivation for building neutrino telescopes is not limited to the observation of AGN. More than 6 orders of magnitude in photon energy, or wavelength, are left unexplored between GeV photons and EeV protons in Fig.~1. Neutrino telescopes have been conceived to fill this gap. Given the history of astronomy, it is difficult to imagine that this will be done without making major, and most likely totally surprising, discoveries. The 18 orders of magnitude in wavelength, from radio-waves to GeV gamma rays, are indeed sprinkled with unexpected discoveries. Neutrinos have the further advantage that, unlike photons of TeV energy and beyond, they are not absorbed by interstellar light. They can, in principle, reach us from the edge of the Universe. If, for instance, the sources of the high energy cosmic rays are well beyond the Virgo cluster, the photon window for their exploration closes above 100~TeV.

Detectors may have to reach kilometer scale before finding the accelerators of the highest energy cosmic rays, improving significantly the searches for cold dark matter particles or WIMPs, and to search meaningfully for the cosmic sources of gamma ray bursts.  In order to explore the plausible region of astronomical parameter space for these fascinating cosmic enigmas, a high energy neutrino telescope must contain several thousand optical modules in a volume of order 1~kilometer on a side. Model building suggests that some, and most likely the most exciting, discoveries may be within reach of much smaller detectors with effective telescope area of order $10^4\rm\,m^2$. Experience with small detectors is, in any case, an important intermediate step and early indications are that a kilometer-scale detector could be constructed in 5 years using existing technologies. 

Consisting of several thousand OMs deployed in natural water or ice, even the ultimate scope of these detectors is similar to that of the Superkamiokande experiment, which presented confirmation of the Kamioka solar neutrino results at this meeting. Being optimized for large effective area rather than low threshold (GeV or more, rather than MeV), they are complementary to SuperK. The challenge to deploy the components in an unfriendly environment is, however, considerable. With a price tag which may be as low as a relatively cheap fixed-target experiment at an accelerator, but could be as high as that of a LHC detector, this must be one of the best motivated large-scale scientific endeavors ever.

As for conventional telescopes, at least two are required to cover the sky. As with particle physics collider experiments, it is very advantageous to explore a new frontier with two or more instruments, preferably using different techniques. This goal may be achieved by exploiting the parallel efforts to use natural water and ice as the Cherenkov medium for particle detection.

There has been a heightened level of activity in this field in the last 3 years. In the Spring of 1993, using the frozen ice as a platform for easy deployment, the Lake Baikal group deployed a small telescope consisting of 36 optical modules. They plan to complete the detector consisting of 200 optical modules by 1998. The Russian-German collaboration presented the neutrino-induced, upgoing muon event shown in Fig.~10. With 19 channels reporting on 4 strings, it is gold-plated. Deep ocean water should be superior in optical quality to that in Lake Baikal. Two collaborations, ANTARES (France) and NESTOR (Greece), are developing the infrastructure and technologies for the deployment of neutrino telescopes in the Mediterranean basin. It is, at this point, only fitting to recall the pioneering role played by the now defunct DUMAND experiment in Hawaii. They made key conceptual and technological contributions to this field.

\begin{figure}[t]
\centering
\hspace{0in}\epsfxsize=2in\epsffile{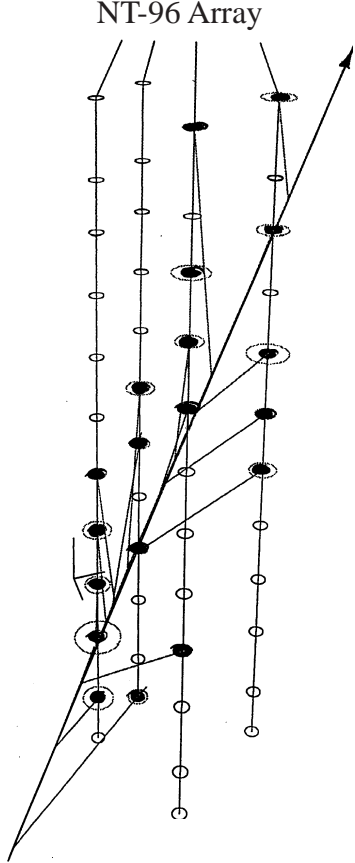}

\caption{An upgoing muon initiated by a neutrino interaction below the Lake Baikal detector. The Cherenkov signal is mapped by 19 optical modules.}
\end{figure}

Using natural Antarctic ice as a particle detector, the AMANDA A detector of 80 OMs, positioned near 1~kilometer depth, has been taking data for almost 3 years. A deep array of $\sim$~300~OMs, AMANDA B, was completed a few days after this conference. In the next season the first strings of kilometer length will be deployed, with the goal to commission a kilometer-scale detector over a period of approximately 5 years. The problem of reconstructing muons in a kilometer-scale detector has been assessed experimentally by i) studying muon tracks registered in both the 1 and 2 kilometer-deep detectors and ii) investigating linear energy response in AMANDA~A.

\renewcommand{\thefigure}{\arabic{figure}a}
\begin{figure}
\centering
\hspace{0in}\epsfysize=6in\epsffile{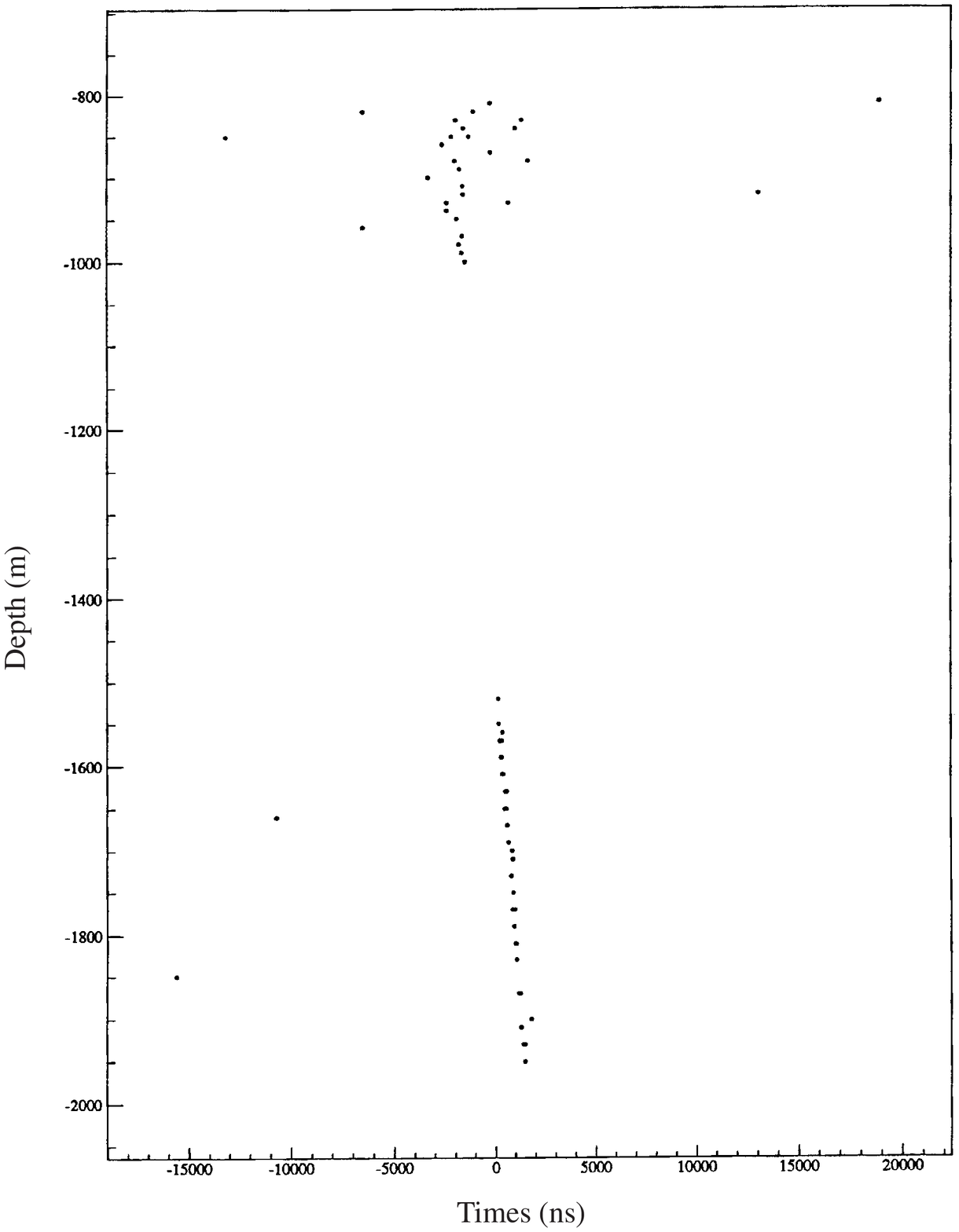}

\caption{Cosmic ray muon track triggered by both AMANDA A and B. Trigger times of the optical modules are shown as a function of depth. The diagram shows the diffusion of the track by bubbles above 1~km depth. Early and late hits, not associated with the track, are photomultiplier noise.}
\end{figure}

\addtocounter{figure}{-1}\renewcommand{\thefigure}{\arabic{figure}b}
\begin{figure}
\centering
\hspace{0in}\epsfysize=6in\epsffile{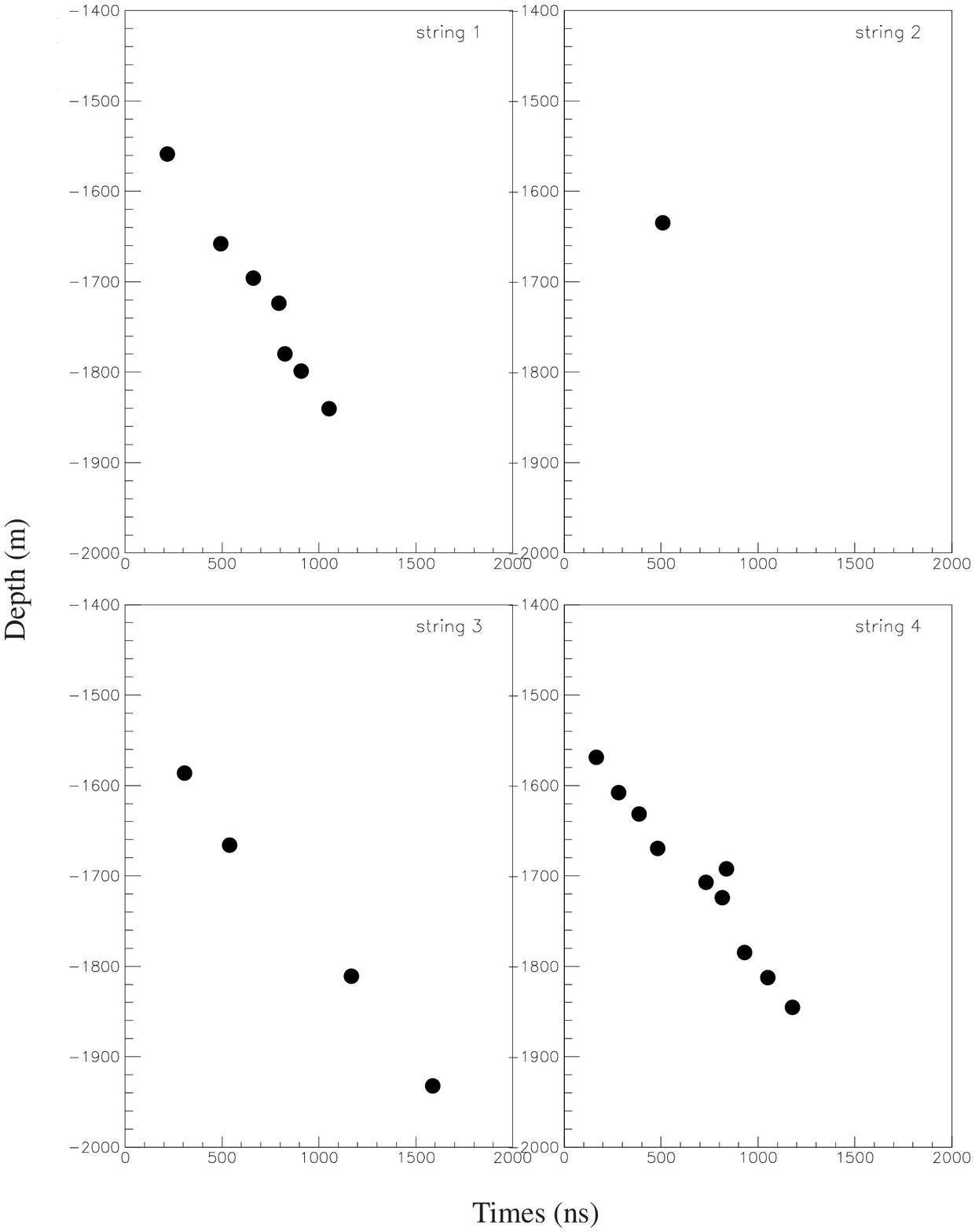}

\caption{Cosmic ray muon track triggered by both AMANDA A and B. Trigger times are shown separately for each string in the deep detector. In this event the muon mostly triggers OMs on strings 1 and 4 which are separated by 79.5~m. }
\end{figure}

Coincidences between AMANDA A and B are triggered at a rate of 0.1~Hz. Every 10 seconds a muon is tracked over 1.2 kilometer; a typical event is shown in Fig.~11. Below 1500~m the vertical muon triggers 2 strings separated by 79.5~m. The distance along the Cherenkov cone is over 100~m, yet, despite some evidence of scattering, propagation of the muon at the speed of light can be readily identified. This, and several other calibration methods indicate a scattering length 2 orders of magnitude larger in the deep detector than in the shallow one. The collaboration has analysed over $5 \times 10^5$  A-B coincidences with no evidence for any source of misreconstructed background events. Reconstruction of such tracks to a degree in the bubble-free ice should not represent a challenge.

Remnant bubbles, much larger in size than expected, limit the scattering length to tens of centimeters in the shallow detector, preventing the reconstruction of muon tracks. With this set-back nature did however provide the AMANDA collaboration with the hint that scattering can be exploited to measure energy. Bubbles very effectively diffuse and contain the light from showers, e.g. initiated by electron neutrinos. The shower can be mapped and its energy measured provided the point of origin is within 50~m of the instrumented ice. Candidate atmospheric neutrino events have been identified in the range of 100~GeV to 1~PeV. A measurement of the spectrum requires a detailed understanding of all systematics. AMANDA~A is the ideal prototype to develop the methods for measuring energy. It is, with its reduced detector dimension and scattering length, a scale model of a kilometer-cubed detector in bubble-free~ice.

Let me conclude by trying to infuse some sanity in the non-debate on ``water and ice". It is a non-debate because, ideally, we want both. Given the pioneering and exploratory nature of the research, we most likely {\bf need} both. All indications are that water and ice have complementary optical properties: while the ``attenuation" lengths are comparable for the blue wavelength photons relevant to the experiments, attenuation is dominated by scattering in ice and by absorption in water. Both have a problem: scattering in ice, potassium decay and bioluminescence in water. Both problems can be solved. The high rates of noise events, especially in a kilometer-size water detector, can be removed by only triggering on pairs of OMs. The solution to the scattering problem in ice has been demonstrated by Monte Carlo simulation. It will be tested on the data which are collected at a rate of $\sim$100~Hz while I am writing this. Because of the fantastic transparency of ice, $\sim$20~OMs report in a typical event, compared to 6$\sim$10 in a water detector with the same effective area. The first photon reaching 5, sometimes more, of the 20 OMs will not be scattered. Their arrival times and the information from the other OMs are merged in a likelihood fit which delivers track reconstruction to better than 2 degrees.

A proposal for a surface detector Hanul, which means ``sky" in Korean, introduces new technology in this neutrino race. It will be composed of modules, somewhat larger than Macro and LVD, which measure muon charge and energy with a large magnet sandwiched between rpc's and Cherenkov counters.

With SuperK, Macro, AMANDA and Baikal this promises to be a happy $\nu$-year.

\section*{Acknowledgements}

The hospitality of Jean Tran Thanh Van and his staff is legendary and has improved, like the finest wine, by over 30 years of aging. We celebrated 30 years of Moriond meetings and all Tran got from me was an AMANDA tee-shirt. On a personal note, I met Vernon Barger at the 1970 meeting in Meribel. He hauled me off to the US and to many great adventures in physics. Thanks Van.

This work was supported in part by the University of Wisconsin Research Committee with funds granted by the Wisconsin Alumni Research Foundation, and in part by the U.S.~Department of Energy under Grant No.~DE-FG02-95ER40896.

\end{document}